\documentclass{article}
\usepackage{spconf,amsmath,graphicx,hyperref}
\usepackage{tcolorbox}
\usepackage{booktabs}
\usepackage{tabularx}
\usepackage{array}
\usepackage{graphicx}
\ninept
\title{CLASH: Counterfactual Auditing of Lexical and Prosodic Reliance in Spoken Sarcasm Detection}
\name{%
\begin{tabular}{@{}c@{}}
Qiyang Sun$^{1,*}$ \quad Xudong Li$^{2,*}$ \quad
Yupei Li$^{1,3}$ \quad Jiabin Xue$^{4}$ \\[3pt]
Yuhang Dai$^{5}$ \quad Jiaming Li$^{6}$ \quad
Bj{\"o}rn W.~Schuller$^{1,3}$
\end{tabular}%
\thanks{$^{*}$These authors contributed equally.}
\thanks{
Corresponding author: Jiaming Li.}
}

\address{%
$^{1}$Imperial College London \quad
$^{2}$New York University 
$^{3}$Technical University of Munich \quad \\
$^{4}$Tencent Inc.
$^{5}$Wuhan University \quad
$^{6}$Nankai University
}

\begin{document}
%\ninept
%
\maketitle
\begin{abstract}
Spoken sarcasm detectors may exploit lexical content, prosody, or their interaction, yet conventional evaluation cannot reveal which cues drive their predictions. We introduce CLASH (Controlled Lexical-Acoustic Separation Harness),
a bilingual counterfactual diagnostic framework that evaluates each utterance under original, lexical-preserving, prosody-preserving, and approximately neutralised conditions. We evaluate handcrafted acoustic-feature systems, self-supervised learning (SSL) probes, and large audio language models (LALMs) on CMMA and MUStARD. For target-only Qwen3-Omni, lexical-preserving speech retains a 0.135--0.148 AUROC advantage over prosody-preserving speech after duration balancing, with cluster-bootstrap intervals above zero; alternative lexical resynthesis preserves this advantage. Acoustic interventions shift scores without consistently improving discrimination or changing binary predictions under the evaluated conditions. Context and interaction estimates vary across corpora. These findings distinguish acoustic sensitivity from sarcasm discrimination while exposing duration, identity, and transformation effects.
\end{abstract}
\section{Introduction}
\label{sec:introduction}

As a task in affective computing \cite{sun2026towards}, sarcasm detection requires recognising intended meanings that differ from, and often oppose, the literal content of an utterance \cite{santosa2025sarcasm}. In speech, this intention may be conveyed through lexical choice, discourse context, vocal delivery, or a combination of these sources \cite{li2025modeling}. Acoustic studies associate sarcasm with variation in fundamental frequency, intensity, voice quality, and speaking rate \cite{rockwell2000lower,caucci2024s}. Such associations do not imply that prosody provides a stable or sufficient decision signal. In spontaneous speech, acoustic realisations of irony vary considerably across utterances and speakers \cite{bryant2010prosodic}. Human judgements also change with discourse context \cite{jang2024context}, and prosodic marking can weaken when lexical semantics already make sarcastic intent salient \cite{li2024functional}. A correct prediction alone therefore does not reveal which cues the detector uses. We address this question by selectively altering lexical content and prosodic delivery and examining the resulting changes in model scores, sarcasm discrimination, and binary predictions.

Automatic spoken sarcasm detection progresses from classifiers based on prosodic, spectral, and contextual descriptors \cite{gao2025spoken} to transfer learning with learned acoustic representations \cite{gao2022deep}. Multimodal corpora such as MUStARD \cite{castro2019towards}, and CMMA \cite{zhang2023cmma} further enable the study of sarcasm in English and Mandarin conversations. Most subsequent work pursues higher predictive performance through feature fusion or incongruity modelling \cite{pan2020modeling}. This direction is useful, but aggregate improvements obtained from audio do not establish that a system uses prosody. Learned speech embeddings can encode lexical and prosodic information within the same representation. Comparing classifiers with and without these embeddings measures the aggregate contribution of speech features, but does not isolate lexical, prosodic, or interaction effects. Likewise, combining information sources does not establish whether their interaction contributes to prediction \cite{hessel2020does}.

Recent diagnostic studies begin to expose this distinction. Prosodically neutral speech synthesis is used to reveal linguistic sensitivity in speech emotion recognition systems \cite{triantafyllopoulos2022probing}. LISTEN evaluates lexical and acoustic reliance in audio language models and reports substantial lexical dominance in emotion recognition \cite{chen2026audio}. For sarcasm, CHARM combines prompt calibration with acoustic evidence to improve detection \cite{sun2026charm}. Recent work identifies prosodic heuristics through modality comparisons and targeted pitch and pause manipulations \cite{chen2026models}. We complement this analysis with a paired factorial design that introduces a neutralised reference and distinguishes score shifts, their label-discriminative AUROC, and binary predictions. This reveals acoustic responses hidden by unchanged decisions. Duration balancing, within-identity comparisons, and alternative lexical resynthesis further establish which intervention contrasts remain robust. These diagnostics address questions left open by the reported heuristic analysis: whether score changes carry label information and whether observed advantages survive duration adjustment and alternative resynthesis, extending paired contrastive evaluation \cite{gardner2020evaluating}.

\begin{figure*}[t]
\centering
\includegraphics[width=0.9\textwidth]{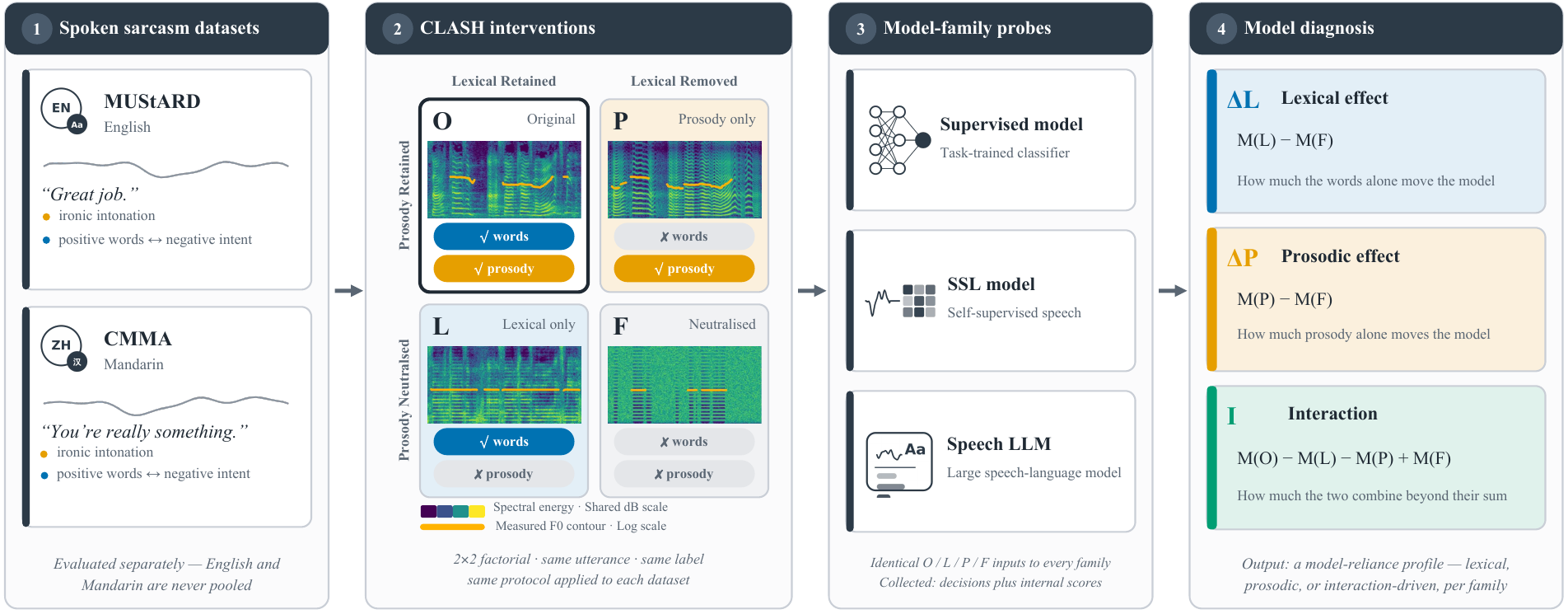}
\caption{Overview of CLASH. Utterances from MUStARD and CMMA undergo paired interventions to form original ($O$), lexical-preserving ($L$), prosody-preserving ($P$), and approximately neutralised ($F$) conditions. The same inputs are evaluated using handcrafted acoustic-feature classifiers, SSL probes, and speech LLMs. Contrasts between conditions characterise lexical and prosodic effects and their interaction. The two corpora are evaluated separately.}
\label{fig:clash_overview}
\vspace{-1em}
\end{figure*}

We therefore introduce CLASH, the Controlled Lexical-Acoustic Separation Harness, a counterfactual diagnostic framework for spoken sarcasm detection. We evaluate handcrafted acoustic-feature systems, self-supervised learning (SSL) probes, and large audio language models (LALMs) on MUStARD and CMMA, representing English and Mandarin, respectively. Context comparisons, alternative text-to-speech resynthesis, and duration balancing assess robustness. For target-only Qwen3-Omni, lexical-preserving speech retains a 0.135--0.148 AUROC advantage over prosody-preserving speech after duration balancing; this advantage also persists under alternative resynthesis.

Our contributions are threefold. First, we introduce an utterance-paired factorial protocol for auditing lexical and prosodic reliance under fixed detector parameters. Second, we demonstrate that acoustic score shifts can coexist with weak label discrimination and unchanged binary predictions. Third, we examine attribution robustness through duration balancing, within-identity comparisons, cluster-bootstrap inference, and alternative lexical resynthesis. The CLASH implementation is available at \url{https://github.com/glam-imperial/clash}.

\section{Methodology}
\label{sec:method}

\subsection{Paired Speech Interventions}

CLASH combines paired speech interventions with complementary measures of score sensitivity, sarcasm discrimination, and binary predictions (Fig.~\ref{fig:clash_overview}). Each utterance has four conditions: \emph{Original} ($O$), \emph{Lexical-preserving} ($L$), \emph{Prosody-preserving} ($P$), and \emph{Approximately neutralised} ($F$). Detector parameters remain fixed across conditions, so within-detector contrasts measure responses to input interventions; cross-detector comparisons additionally reflect representation and readout differences. The original label measures discrimination retained after transformation, without assuming that perceived sarcasm remains unchanged.

All detector inputs use 16\,kHz mono audio. For condition $L$, voiced-frame F0 is set to the utterance median in English. In Mandarin, deviations from the utterance median log-F0 are scaled by 0.5, reducing pitch variation while retaining lexical tone-contour shapes. Non-silent frames are normalised to their median RMS energy, with timing and pauses unchanged. Thus, $L$ attenuates pitch and intensity variation without eliminating rhythmic cues. Condition $P$ uses WORLD resynthesis \cite{morise2016world}, replacing the original time-varying spectral envelope with a fixed carrier envelope while retaining F0 and aperiodicity parameters. This replacement disrupts segmental spectral cues to reduce lexical intelligibility. Condition $F$ combines these operations; retained timing and processing effects preclude an information-free baseline.

To test whether the lexical-preservation advantage depends on WORLD-based processing, we construct an alternative $L$ condition, denoted $L_{\mathrm{TTS}}$, by synthesising each utterance's reference transcript with Qwen3-TTS-12Hz-1.7B-CustomVoice \cite{hu2026qwen3}, using a fixed voice and neutral-delivery instructions. This supplementary condition leaves the original $O/L/P/F$ design unchanged. Whisper-based word/character error rates, F0 and energy correlations, and ECAPA-TDNN \cite{desplanques2020ecapa} similarities assess intelligibility, acoustic preservation, and speaker changes across the generated conditions.

\subsection{Detector Evaluation}

We compare handcrafted acoustic-feature systems, SSL probes based on WavLM \cite{chen2022wavlm} and wav2vec~2.0 \cite{baevski2020wav2vec}, and LALMs. Main estimates use 3,960 CMMA and 690 MUStARD utterances; additional cross-model comparisons use a common held-out subset of 781 and 163 utterances, respectively. These evaluations remain separate. The additional comparison includes WavLM-Base-Plus, wav2vec~2.0-Base, Qwen2.5-Omni-7B, Qwen2-Audio-7B, and Voxtral-Mini-3B \cite{liu2025voxtral}.

We compare handcrafted and learned acoustic representations under the same classifier and training protocol. The handcrafted baseline uses 88-dimensional eGeMAPSv02 functionals extracted with openSMILE \cite{eyben2015geneva}, providing explicit prosodic, spectral, and voice-quality descriptors. The learned baseline uses frozen WavLM-Large \cite{chen2022wavlm} (\texttt{microsoft/wavlm-large}), with temporal mean pooling of the final hidden layer. This comparison examines whether intervention responses depend on the representation, without assuming that either system selectively encodes lexical or prosodic information. Both systems use standardised features and L2-regularised logistic regression trained on original CMMA training speech only; three-fold training-set cross-validation selects regularisation by AUROC. MUStARD therefore assesses cross-lingual, cross-corpus transfer for these probes. Auxiliary audits compare logistic regression, linear regression, XGBoost, and LightGBM heads and examine train--test feature separation. The main LALM is Qwen3-Omni-30B-A3B-Instruct \cite{xu2025qwen3}, evaluated without fine-tuning using greedy decoding and the fixed instruction below.

\begin{tcolorbox}[colback=gray!5, colframe=gray!50, title=Fixed Detection Prompt, fonttitle=\small\bfseries, boxrule=0.4pt, arc=1mm, left=1mm, right=1mm, top=1mm, bottom=1mm]
\small
Determine whether the speaker is sarcastic. Answer with exactly one word: Yes or No.
\end{tcolorbox}

The context-aware setting prepends preceding dialogue text to the target-only input. Context remains identical across speech conditions, so context-aware $F$ retains contextual evidence.

\subsection{Diagnostic Contrasts and Robustness}

Let $s_i^c$ denote the continuous score for utterance $i$ under condition $c\in\{O,L,P,F,L_{\mathrm{TTS}}\}$, and let $A_c$ denote its condition-level AUROC against the original sarcasm labels. Scores use logistic regression decision values or LALM affirmative-versus-negative token log-probability masses. Paired differences, $d_i=s_i^P-s_i^F$, measure score sensitivity. The AUROC of $d_i$ assesses whether these changes distinguish sarcastic from non-sarcastic utterances; it differs from $A_P-A_F$, which compares discrimination under the two conditions. Macro-F1 and positive-prediction rates describe binary predictions at a fixed score threshold of zero. We summarise standardised score sensitivity within each context setting as $d_z=\overline{d}/\operatorname{sd}(d)$.

Principal contrasts are $A_L-A_P$ and $A_{L_{\mathrm{TTS}}}-A_P$; supplementary contrasts are
\begin{equation}
\Delta_L=A_L-A_F,\quad \Delta_P=A_P-A_F,
\end{equation}
\begin{equation}
I=A_O-A_L-A_P+A_F.
\end{equation}
These contrasts quantify discrimination under the implemented transformations for each detector. Although $\Delta_L-\Delta_P=A_L-A_P$ removes dependence on $A_F$, off-target differences between $L$ and $P$ remain uncontrolled. The interaction $I$ describes AUROC non-additivity, not a specific incongruity mechanism.

For context comparisons, we define $D=\Delta_{P,\mathrm{target}}-\Delta_{P,\mathrm{context}}$. Within-identity AUROC restricts positive--negative comparisons to the same conversation on CMMA or speaker on MUStARD. A separate processing-sensitivity check uses the detector scores to distinguish $O$ from $F$, reporting $\max(A,1-A)$ to allow either score orientation.

To assess sensitivity to duration, we partition each corpus into five quantile bins of $\log(1+t_i)$, where $t_i$ denotes the original duration of utterance $i$. Within each bin, weighting equalises the total contribution of the two classes:
\begin{equation}
w_i = \frac{\min\!\left\{n_{b(i),0},\,n_{b(i),1}\right\}}{n_{b(i),y_i}},
\label{eq:duration_weight}
\end{equation}
where $b(i)$ denotes the bin containing utterance $i$, $y_i\in\{0,1\}$ is its sarcasm label, and $n_{b,y}$ is the number of utterances with label $y$ in bin $b$. The same weights apply to all speech conditions when computing weighted AUROC. A ten-bin analysis assesses sensitivity to binning. We estimate 95\% percentile confidence intervals using 2,000 cluster-bootstrap replicates \cite{field2007bootstrapping}, retaining all paired speech conditions and context settings and recomputing the contrasts and weights within each replicate. Bin boundaries remain fixed. Resampling uses conversation clusters for CMMA and 21 speaker clusters for MUStARD, where reliable conversation identifiers are unavailable.

\begin{table*}[t]
\centering
\caption{Performance on CMMA and MUStARD, unadjusted for duration. Cells report AUROC / Macro-F1; bold marks the highest value for each metric within each row. The final column reports the AUROC contrast $A_L-A_P$. $L_{\mathrm{TTS}}$ provides an alternative to WORLD-based lexical preservation. LR denotes logistic regression. Preceding dialogue text remains available across all conditions in context-aware evaluation.}
\label{tab:main_results}
\begingroup
\small
\setlength{\tabcolsep}{3pt}
\renewcommand{\arraystretch}{1.08}
\begin{tabularx}{\textwidth}{@{}l l *{5}{>{\centering\arraybackslash}X} r@{}}
\toprule
System & Corpus & $O$ & $L$ & $P$ & $F$ & $L_{\mathrm{TTS}}$ & $A_L-A_P$ \\
\midrule
\multicolumn{8}{@{}l}{\textit{Target-only evaluation}} \\
eGeMAPS + LR & CMMA & .466 / .471 & .511 / .474 & .513 / \textbf{.502} & .521 / .489 & \textbf{.531} / .496 & $-.002$ \\
eGeMAPS + LR & MUStARD & \textbf{.554} / .423 & .543 / .390 & .522 / \textbf{.464} & .456 / .445 & .499 / .336 & $+.021$ \\
WavLM-Large + LR & CMMA & \textbf{.584} / \textbf{.490} & .521 / .472 & .499 / .470 & .502 / .470 & .523 / .472 & $+.022$ \\
WavLM-Large + LR & MUStARD & \textbf{.612} / \textbf{.370} & .571 / .361 & .568 / .332 & .442 / .333 & .524 / .333 & $+.003$ \\
Qwen3-Omni & CMMA & \textbf{.734} / \textbf{.602} & .688 / .594 & .549 / .470 & .471 / .470 & .695 / .557 & $+.139$ \\
Qwen3-Omni & MUStARD & \textbf{.775} / \textbf{.690} & .669 / .619 & .370 / .333 & .343 / .333 & .687 / .625 & $+.299$ \\
\midrule
\multicolumn{8}{@{}l}{\textit{Context-aware evaluation}} \\
Qwen3-Omni & CMMA & \textbf{.774} / \textbf{.615} & .756 / .610 & .682 / .565 & .673 / .571 & .766 / .558 & $+.074$ \\
Qwen3-Omni & MUStARD & \textbf{.780} / \textbf{.658} & .682 / .583 & .499 / .488 & .486 / .477 & .707 / .634 & $+.183$ \\
\bottomrule
\end{tabularx}
\endgroup
\vspace{-1em}
\end{table*}

\section{Experiments and Results}
\label{sec:experiments}
\subsection{Cross-Model Intervention Profiles}

Table~\ref{tab:main_results} reports AUROC and Macro-F1 on CMMA and MUStARD, whose sarcasm prevalence is 11.4\% and 50.0\%, respectively.

Qwen3-Omni achieves the highest original-speech AUROC among the systems in Table~\ref{tab:main_results}, and its lexical-preserving condition retains substantial discrimination. The unadjusted $A_L-A_P$ contrast is 0.139 on CMMA and 0.299 on MUStARD. The acoustic and SSL probes show different patterns. eGeMAPS provides weak discrimination across conditions, whereas WavLM-Large performs better on original speech but exhibits little separation between $L$ and $P$, particularly on MUStARD. Thus, stronger original-speech performance does not consistently correspond to a larger prosody-preservation advantage.

These representation baselines expose limitations of probe-based attribution. Across four classifier heads, original-speech CMMA AUROC ranges from 0.456 to 0.503 for eGeMAPS and from 0.552 to 0.610 for WavLM-Large. Train--test domain classification yields 0.810/0.848 AUROC, indicating distributional separation. Their detector scores distinguish $O$ from $F$ at 0.824/0.818 AUROC. This separation can reflect intended cue changes or processing artefacts; it does not identify their relative contributions. These profiles therefore characterise the tested representations, readouts, and transfer settings rather than general properties of model families.

\begin{table}[t]
\centering
\caption{Additional model comparisons on the common CMMA held-out subset. Here, $\Delta_L=A_L-A_F$ and $\Delta_P=A_P-A_F$. The final column gives $\Delta_L-\Delta_P=A_L-A_P$. Estimates are unadjusted and reported to three decimal places.}
\label{tab:additional_models}
\begingroup
\setlength{\tabcolsep}{3pt}
\renewcommand{\arraystretch}{1.08}
\begin{tabularx}{\columnwidth}{@{}Xrrr@{}}
\toprule
System & $\Delta_L$ & $\Delta_P$ & $A_L-A_P$ \\
\midrule
eGeMAPS + LR       & $-.027$ & $-.085$ & $.058$ \\
WavLM-Base-Plus    & $-.001$ & $-.048$ & $.047$ \\
wav2vec~2.0-Base   & $.001$  & $-.024$ & $.025$ \\
\midrule
Qwen2.5-Omni-7B    & $.258$  & $.039$  & $.219$ \\
Qwen2-Audio-7B     & $.056$  & $-.033$ & $.089$ \\
Voxtral-Mini-3B    & $.050$  & $-.021$ & $.071$ \\
\bottomrule
\end{tabularx}
\vspace{-1em}
\endgroup
\end{table}

Table~\ref{tab:additional_models} extends the detector comparison on a common CMMA subset ($n=781$). All six systems yield positive $A_L-A_P$ contrasts, ranging from 0.025 to 0.058 for acoustic probes and from 0.071 to 0.219 for LALMs. However, negative $\Delta_P$ values for the probes show that a relative lexical-preservation advantage need not imply strong absolute discrimination. These unadjusted CMMA results extend the contrast direction beyond Qwen3-Omni; the full robustness assessment remains model-specific.

\subsection{Score Sensitivity and Binary Predictions}

Paired score diagnostics distinguish score sensitivity, label discrimination, and binary predictions. Under target-only Qwen3-Omni, $P$ increases log-odds relative to $F$ for 74.8\% of CMMA utterances and 82.5\% of MUStARD utterances, yet both conditions produce exclusively non-sarcastic predictions. The paired differences yield AUROC values of 0.566 and 0.524, respectively. WavLM-Large shows a similar dissociation on CMMA, with a mean shift of 0.567 but a difference-based AUROC of 0.502; on MUStARD, the corresponding AUROC is 0.596. The same paired inputs therefore yield different conclusions at the score, discrimination, and decision levels. Acoustic responsiveness alone is consequently an insufficient criterion for evaluating effective prosodic cue use in the tested sarcasm detectors.

\begin{figure}[t]
\centering
\includegraphics[width=0.48\textwidth]{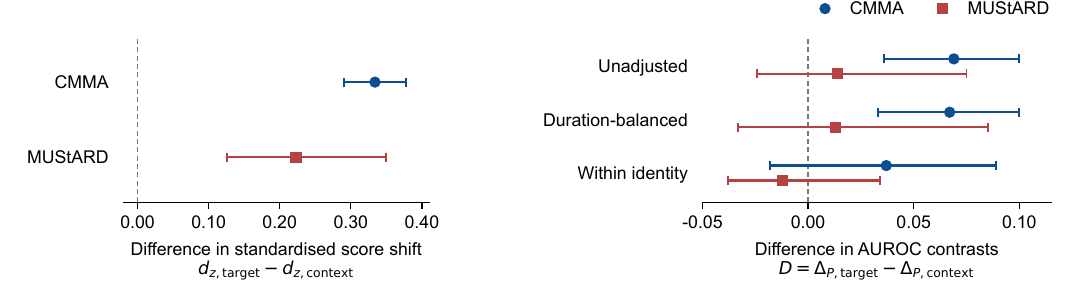}
\caption{Qwen3-Omni context comparison. Left: target-only minus context-aware standardised $P-F$ score shifts. Right: $D=\Delta_{P,\mathrm{target}}-\Delta_{P,\mathrm{context}}$. Within-identity comparisons use CMMA conversations or MUStARD speakers. Bars show marginal 95\% paired cluster-bootstrap intervals; dashed lines mark zero. The axes measure different quantities.}
\label{fig:context_diagnosis}
\vspace{-1em}
\end{figure}

\subsection{Duration and Lexical Resynthesis}

Because the interventions retain temporal structure, we examine whether duration provides a residual label-related cue that affects the intervention contrasts. On MUStARD, duration alone yields an AUROC of 0.660, and sarcastic utterances average 5.77\,s compared with 4.45\,s for non-sarcastic utterances. Qwen3-Omni scores on unintelligible speech correlate negatively with duration ($\rho=-0.72$). Longer sarcastic utterances can therefore receive lower scores after lexical removal, contributing to the below-chance ordering in $P$ and $F$.

After duration balancing, $A_F$ increases from 0.343 to 0.447 on MUStARD, while $A_L-A_P$ decreases from 0.299 to 0.135. Its direction remains positive. CMMA changes less, with the corresponding contrast increasing from 0.139 to 0.148. Table~\ref{tab:robustness} shows that both duration-balanced contrasts have cluster-bootstrap intervals above zero.

\begin{table}[t]
\centering
\caption{Duration-balanced AUROC contrasts for target-only Qwen3-Omni. Intervals are marginal 95\% cluster-bootstrap confidence intervals, using conversations for CMMA and speakers for MUStARD.}
\label{tab:robustness}
\begingroup
\small
\setlength{\tabcolsep}{2pt}
\renewcommand{\arraystretch}{1.08}
\begin{tabularx}{\columnwidth}{@{}l c >{\centering\arraybackslash}X c@{}}
\toprule
Corpus & Contrast & Estimate & 95\% CI \\
\midrule
CMMA & $A_L-A_P$ & .148 & [.103, .193] \\
 & $A_{L_{\mathrm{TTS}}}-A_P$ & .157 & [.113, .202] \\
MUStARD & $A_L-A_P$ & .135 & [.094, .210] \\
 & $A_{L_{\mathrm{TTS}}}-A_P$ & .173 & [.118, .276] \\
\bottomrule
\end{tabularx}
\endgroup
\vspace{-1em}
\end{table}

Target-only Qwen3-Omni retains positive duration-balanced lexical-preservation contrasts under WORLD and TTS, with intervals above zero (Table~\ref{tab:robustness}). This agreement concerns AUROC; binary decisions differ. On CMMA, TTS raises positive predictions from 14.4\% to 30.8\%, while Macro-F1 falls from 0.594 to 0.557. Robustness is model-specific: on MUStARD, eGeMAPS and WavLM-Large yield $A_{L_{\mathrm{TTS}}}-A_P=-0.023$ and $-0.044$, respectively. Alternative resynthesis supports the Qwen3-Omni contrast without establishing equivalent transformation effects across the generated speech conditions.

Relative to $O$, WORLD-based $L$ increases CMMA CER/MUStARD WER by 19.9/11.0 percentage points; $L_{\mathrm{TTS}}$ reduces them by 30.4/15.4 points. Two experts each inspect all four $O/L/P/F$ versions of 50 utterances per corpus and judge the inspected generated samples to have acceptable perceptual quality. These checks support transformation quality without establishing perfect cue separation.

\subsection{Context Effects across Corpora}

For Qwen3-Omni, we examine whether removing dialogue context increases the discriminative benefit of $P$ relative to $F$. Target-only inputs yield larger standardised $P-F$ score shifts than context-aware inputs: $d_z$ is 0.77 versus 0.44 on CMMA and 1.02 versus 0.80 on MUStARD. This sensitivity does not establish greater prosodic discrimination (Fig.~\ref{fig:context_diagnosis}). On CMMA, $D$ is 0.069 [0.036, 0.100] before adjustment and 0.067 [0.033, 0.100] after duration balancing, but 0.037 [$-0.018$, 0.089] within conversations. The within-speaker MUStARD estimate is $-0.012$ [$-0.038$, 0.034]. Both within-identity intervals include zero, providing insufficient evidence for a consistent compensatory benefit from removing context.

The additional value of original delivery also varies. Within-conversation analysis reduces the target-only $A_O-A_L$ contrast on CMMA from 0.046 to 0.006, whereas the within-speaker MUStARD contrast remains 0.090, compared with 0.106 overall. After duration balancing, $I$ is 0.092 on MUStARD (95\% CI [0.014, 0.146]) and $-0.030$ on CMMA. The recurring lexical-preservation advantage therefore coexists with corpus-dependent context and joint-condition effects. Since language and corpus vary together, these observations do not isolate cultural mechanisms.

\section{Conclusion}
\label{sec:conclusion}
We introduced CLASH, a paired diagnostic framework for examining lexical and prosodic cue reliance in spoken sarcasm detection. Across CMMA and MUStARD, the systems exhibited response patterns, and acoustic interventions shifted model scores without consistently improving sarcasm discrimination or changing binary predictions. For target-only Qwen3-Omni, the lexical-preservation advantage persisted after duration balancing and under independent lexical resynthesis, whereas context and interaction effects varied across corpora. These findings distinguished acoustic sensitivity from discrimination and showed why audio-based performance gains alone provided insufficient evidence of prosodic cue use. Imperfect cue separation, probe transfer, and corpus-specific structure limited the scope of the attribution.

\newpage

\bibliographystyle{IEEEbib}
\bibliography{strings,refs}

\end{document}